\documentclass[12pt]{article}

\usepackage{epsfig}
\usepackage{amsmath}
\usepackage{soul,color}
\usepackage{bm}
\usepackage{epsfig}
\usepackage{hyperref}
\usepackage{multirow}
\usepackage{latexsym, graphicx, alltt}
\usepackage{tikz} 
\usetikzlibrary{calc,shapes} 
\usetikzlibrary{arrows,snakes,backgrounds}
\tikzstyle{arr}=[-latex,black]
\tikzstyle{latent}=[circle,draw,inner sep=0pt,minimum size=6mm]
\tikzstyle{manifest}=[rectangle,draw,inner sep=0pt,minimum size=6mm]
\usepackage{float}
\usepackage[nolists]{endfloat}

\newcommand{\tr}{^{\prime}}
\def\b#1{\mbox{\boldmath $#1$}}    

\newcommand{\la}{\lambda}

\date{}

\begin{document}

\title{Evaluation of student proficiency through a multidimensional finite mixture IRT model}

 \author{Silvia Bacci\footnote{Department of Economics, 
University of Perugia (IT).}  \and
Francesco Bartolucci$^*$ \and
Leonardo Grilli\footnote{Department of Statistics, Computer Science, Applications ``G. Parenti'', University of Florence (IT).} \and
Carla Rampichini$^{\dag}$ 
}

\maketitle

\def\baselinestretch{1.2}
\begin{abstract}
In certain academic systems, a student can enroll for an exam immediately after the end of the teaching period or can postpone it to any later examination session, so that the grade is missing until the exam is not attempted. We propose an approach for the evaluation {\em in itinere} of a student's proficiency accounting also for non-attempted exams. The approach is based on considering each exam as an item, so that responding to the item amounts to attempting the exam, and on an Item Response Theory model that includes two latent variables corresponding to the student's ability and the propensity to attempt the exam. In this way, we explicitly account for non-ignorable missing observations as the indicators of item response also contribute to measure the ability. 
The two latent variables are assumed to have a discrete distribution defining latent classes of students that are homogeneous in terms of ability and priority assigned to exams. The model, which also allows for individual covariates in its structural part, is fitted by the Expectation-Maximization algorithm. The approach is illustrated through the analysis of data about the first-year exams of freshmen of the School of Economics at the University of Florence (Italy).\\


\end{abstract}

\newpage

\section{Introduction} \label{sec:1}

The analysis of university student careers is relevant for both planning and guidance. In particular, the proficiency of students at the end of the first academic year  is highly predictive of the final outcome; its evaluation can be based on the total number of gained credits (Grilli, 2016)  or on the result   at each compulsory exam. Each exam can be passed or failed; in the Italian university system, for instance, a passed exam is evaluated on a scale ranging from 18 to 30, plus ``30 with honors''. 
In this regard, Bertaccini, Grilli, and Rampichini (2013)  proposed an IRT-MIMIC model where each compulsory first-year exam corresponds to a binary item, with the success standing for having passed the exam within the first year.  
However, the proposed model does not account for two aspects which are common in the Italian university system: (i) courses with a large number of students are divided into parallel groups; (ii) a student can take the compulsory exams in any order and not necessarily during the first year. 
The first aspect may be addressed by considering a distinct item  for each exam group. The second aspect requires to extend the model to account for the student strategy in choosing to take an exam during the first year or to postpone it. Thus, for a certain student at a given time point the result of a certain exam can be missing for two reasons: (i) the  item corresponding to that exam is not due since the student belongs to another group; (ii) the exam is due, but the student decided to postpone it. The first kind of missing data is structural and thus it can be assumed to be ignorable, whereas the second kind of missing data is potentially informative, as it could be related to the student's ability, which is measured by the exam. 

In statistical terms, postponing an exam generates missing not-at-random (MNAR) data which are then not ignorable (Little and Rubin, 2002; Mealli and Rubin, 2015). To handle this kind of missingness, we follow the original idea of Lord (1983), further developed in the parametric setting by (Holman and Glas, 2005). In particular,  we assume that the student's performance is driven by two latent variables. The first main latent variable affects both the enrollment decision and the exam result, thus representing student ability which is of main interest; the second latent variable only affects the enrollment decision, thus representing the student's priority in taking the exams.

In this paper, we focus on the analysis of student careers at the University of Florence, considering freshmen of the academic year 2013/2014 who are enrolled in the degree programs \emph{Business} and \emph{Economics}. These programs share the six compulsory courses of the first year. Given the large number of freshmen, each course is organized in four parallel groups on the basis of the first letter of the student's surname.  
This entails  a set of $6 \times 4= 24$ items, thus generating the structural missing values mentioned above.
A student can take exams in any order during the examination sessions of the academic year (January, February, June, July, September, December).  To take a certain exam in an examination session, the student has to enroll via a web procedure, which is also used to record the exam result.  Most freshmen cannot manage the entire workload, so they decide to postpone one or more exams to the following academic year. A student postponing an exam never enrolls for that exam during the first year, thus generating a missing value that likely is informative.

The main purposes of our analysis are: (i) evaluating the student performance on the basis of the decision to take or postpone each of the compulsory first-year exams, in addition to the grades of passed exams;
(ii) characterizing the compulsory exams in terms of their difficulty and discrimination power; (iii) comparing the parallel groups of each exam to check whether they behave similarly; (iv) clustering students into homogeneous classes of ability and preference for the exams sequence, controlling for observed student characteristics. The results of our study can help in planning the degree programs and organizing student tutoring. 

To perform the analysis, we develop an Item Response Theory model (IRT; Hambleton and Swaminathan, 1985;  Van der Linden and Hambleton, 1997;  Bartolucci, Bacci, and Gnaldi, 2015)  for multidimensional latent traits (Reckase, 2010) accounting for both structural and non-ignorable missing values. Exams are treated as ordinal items measuring the latent variable representing the student ability, whereas binary indicators of exam enrollment measure both ability and another latent variable representing exam priority. This structure, where  a set of items contributes to measure more latent variables, is known as \emph{within-item multidimensionality} (Adams, Wilson, and Wang, 1997) and the corresponding IRT model with continuous latent traits is known  as \emph{item bifactor model}, which is a special case of the confirmatory item factor analysis model (Gibbons and Hedeker, 1992; Gibbons et al., 2007; Cai, 2010). 
The item bifactor model assumes mutually uncorrelated latent variables, with a general latent trait affecting all items through suitable loadings and other traits for certain specific subsets of items. The model we propose is  more general for two aspects: (i) it allows for certain forms of correlation among latent variables; 
(ii) it is not necessary to have a general latent trait affecting  all items.  
Moreover, we assume that the latent traits have a discrete rather than continuous distribution. This choice increases the flexibility of the model and allows us to cluster individuals in homogeneous Latent Classes (LC; Lazarsfled and Henry, 1968; Goodman, 1974). We let also allow for class membership probability to depend on student covariates.
The  proposed model is an extension of the LC-IRT model of Bacci and Bartolucci, 2015, which accounts for informative missing responses through a similar latent structure, but is limited to binary items and does not admit structural missing values. 
We observe that the application here proposed drives the development of a very general model that is suitable for a wide range of applications in other fields of knowledge, other than the educational setting, involving the measurement of multiple latent traits. 

The procedures to estimate the proposed multidimensional LC-IRT model are implemented in  the {\tt R} package {\tt MLCIRTwithin} (Bartolucci, Bacci, 2015), freely downloadable from \verb' http://CRAN.R-project.org/package=MLCIRTwithin'. 

The remainder of the paper is organized as follows. Section \ref{sec:data} describes the data. Section \ref{sec:mod} illustrates  the model  and  Section \ref{sec:inference} provides details on estimation.  Section \ref{sec:applic} is devoted to the application; in particular, this section  describes model selection, including tests for the ignorability of the missing data mechanism  and for the homogeneity of groups of the same academic course. Section \ref{sec:concl} provides main conclusions.

\section{Data description}\label{sec:data}

The data  set for the analysis is obtained from the administrative archive on student careers, considering the freshmen of the academic year 2013/2014 enrolled in the degree programs \emph{Business} (`Economia Aziendale') and \emph{Economics} (`Economia e Commercio') of the University of Florence. 

The data set includes background characteristics of the students and their careers until December 2014. 
The first year entails six compulsory courses, three in the first semester and three in the second semester.
All courses have parallel classes with distinct teachers, according to the first letter of the student's surname.
Five courses are common to the two degree programs, and are divided in four groups (A-C, D-L, M-P, Q-Z), while the course Management differs between the two degree programs, and the students are split in two groups for each program (A-L, M-Z).

Students can take the exams in any order,  not necessarily at the end of the corresponding course. The exams of the first-semester courses can be taken in any of the six sessions from January to December, while the exams of the second-semester courses can be taken in any of the four sessions from June to December. In order to take the exam in the chosen session, the student has to enroll via web. If a student decides to postpone an exam to the next academic year, the enrollment record for that exam is empty.

In the analysis we consider the $861$ freshmen who enrolled for at least one exam until December 2014 (89\% of the freshmen).
For each student, the data set contains information on the number of enrollments to each of the six exams, alongside with their outcomes. Passed exams are scored with integer values ranging from $18$ to $30$, plus ``$30$ with honors''. 
For each exam (merging groups), Table \ref{tab:courses} reports the percentage of students who enrolled in at least one of the six sessions of 2014 (\emph{enrollment rate}), the distribution of the outcome for students enrolled at least once, considering the best outcome if the exam is repeated, and the percentage of students who passed the exam by December 2014 (\emph{passing rate}), both conditional on enrollment and overall. The \emph{overall passing rate} is obtained as the product of the \emph{enrollment rate} by the \emph{conditional passing rate}.

\begin{table}[!ht]
  \centering
  \caption{\label{tab:courses}\em Enrollment rates and exam results of first-year exams by course. Freshmen 2013/2014, University of Florence, degree programs \emph{ Business} and \emph{Economics}, examination sessions from January to December 2014.}
\small
\centering
 \begin{tabular}{lc|c|c|c|c|c|c|c|c} \hline\hline
 \multicolumn{2}{c|}{Course (semester)} & Enrollment   &\multicolumn{5}{|c|}{Exam grade (\%)}        &\multicolumn{2}{|c}{Passing rate (\%)}\\ \cline{4-10} 
                &       & rate (\%)     &failed      &$18$-$21$&$22$-$24$&$25$-$27$&$\geq28$& enrolled   &overall  \\ 
\hline 
 Accounting     &(I) &93.5&42.5&15.9&17.3&17.0& 7.3 &57.5&53.8  \\
 Mathematics    &(I) &67.8&65.8&16.2& 7.3& 6.8& 4.0 &34.2&21.1  \\
  Law           &(I) &48.3&47.1&14.2&16.1&14.4& 8.2 &52.9&25.6  \\
 Management     &(II)&72.5&30.6& 8.2&16.7&23.2&21.3 &69.4&50.3  \\
 MicroEcon      &(II)&41.8&41.9&10.6&11.4&18.1&18.1 &58.1&24.3  \\
 Statistics     &(II)&67.0&39.7&16.8&13.5&11.4&18.5 &60.3&40.4  \\
\hline                                                                                                                           \end{tabular}
\end{table}

Table \ref{tab:courses} highlights the large variability of student performance across the courses.
It is worth noting that the overall passing rate may result from markedly different patterns. For example, the overall passing rate for Accounting is higher than for Law ($53.8\%$ versus $25.6\%$), which is mainly due to different enrollment rates ($93.5\%$ versus $48.3\%$), whereas the conditional passing rates are similar ($57.5\%$ versus $52.9\%$). On the contrary, the higher overall passing rate for Statistics with respect to Mathematics ($40.4\%$ versus $21.1\%$) is mainly due to different conditional passing rates ($60.3\%$  versus $34.2\%$), while being the enrollment rates similar ($67.0\%$ versus $67.8\%$). 

Table \ref{tab:characteristics} shows the performance of freshmen by gender,  High School (HS) type,   HS grade (ranging from 60 to 100), late matriculation, degree program, and course group. 
The table reports the average number of attempted exams (enrolled at least once) and the number of passed exams.

\begin{table}[!ht]
\caption{\label{tab:characteristics} \em Average number of attempted and passed first-year exams by student characteristics. Freshmen 2013/2014, University of Florence, degree programs \emph{ Business} and \emph{Economics}, examination sessions from January to December 2014.}
\small 
\centering
\begin{tabular}{lrcc} \hline\hline
    \noalign{\smallskip}
    &    \multicolumn{1}{c}{$N$}   & \multicolumn{2}{c}{Average number of exams} \\ \cline{3-4}
& & \multicolumn{1}{c}{enrolled to} & \multicolumn{1}{c}{passed} \\  \hline 
All freshmen                &     861&    3.8& 2.2 \\
\emph{Gender}               &        &        &      \\
\hspace{2 ex} Male          &     502&     3.8& 2.1 \\
\hspace{2 ex} Female        &     359&     3.9& 2.2 \\
\emph{HS type}              &        &        &      \\
\hspace{2 ex} Technical     &     765&     3.8& 2.1 \\
\hspace{2 ex} Humanities    &     201&     3.7& 1.9 \\
\hspace{2 ex} Scientific    &     321&     4.1& 2.4 \\
\hspace{2 ex} Other         &     284&     3.6& 1.8 \\
\emph{HS grade}            &        &        &      \\
\hspace{2 ex} $ < 80$       &     596&     3.6& 1.7 \\
\hspace{2 ex} $ \ge 80$     &     265&     4.4& 3.3 \\
\emph{Late matriculation}               &        &        &      \\
\hspace{2 ex} No            &     759&     4.0& 2.3 \\
\hspace{2 ex} Yes           &     102&     3.0& 1.3 \\
\emph{Degree program}      &        &        &      \\
\hspace{2 ex} Business      &     588&    3.8& 2.1 \\
\hspace{2 ex} Economics     &     273&    3.9& 2.3 \\
\emph{Course group}                &        &        &      \\
\hspace{2 ex} A-C           &     257&   3.7 & 2.2 \\
\hspace{2 ex} D-L           &     240&   3.8 & 2.2 \\
\hspace{2 ex} M-P           &     204&   4.0 & 2.1 \\
\hspace{2 ex} Q-Z           &     160&   3.9 & 2.3 \\
\hline
\end{tabular}
\end{table}

Considering the six compulsory courses, on average students attempted 3.8 exams and passed 2.2 exams, with students from Scientific high schools or with a high HS grade (greater than 80 out of 100) perform better. On the contrary, late matriculated students perform worse in terms of both attempted and passed exams.

\section{Model formulation}\label{sec:mod} 
A distinctive feature of the case study under consideration is represented by missing observations on exam results, which could reflect student ability. Indeed, we expect that the tendency to attempt a certain exam in a given session is higher for students with greater ability so that exam results are not missing at random. 

In general, data are missing at random (MAR) if the conditional distribution of the response indicator, given  observed and unobserved data, is the same whatever the unobserved data for all the parameters values (see Definition 1, Mealli and Rubin, 2015). If this condition does not hold,  data are MNAR (see Definition 1, Mealli and Rubin, 2015),  thus the missingness mechanism  is non-ignorable and it should be explicitly modeled to avoid wrong inferential conclusions.

In the statistical literature, different approaches have been proposed to model  MNAR data, including: (i) the {\em selection approach} (Diggle and Kenward, 1994), in which a model is specified for the marginal distribution of the complete (i.e., observed and unobserved) data  and the conditional distribution of the missing indicators, given these data; (ii) the  {\em pattern-mixture approach} (Little, 1993), in which a model is formulated for the marginal distribution of the missing indicators and the conditional distribution of the complete data, given these indicators; (iii) the {\em shared-parameter approach} (Wu and Carroll, 1988; Follman and Wu, 1995), 
which introduces a latent variable to capture the association between the observed responses and the missing process. An example of a shared-parameter approach in the IRT framework is provided by  the finite mixture Structural Equation Model (SEM) of (Bacci and Bartolucci, 2015); see also Jedidi, Jagpal, and DeSarbo (1997), Dolan and van der Maas (1998), and Arminger, Stein, and Wittenberg (1999) for details on finite mixture SEMs.

In particular, the  model of  Bacci and Bartolucci (2015)  is characterized by a set of multiple equations that define the  
relationships among latent variables and  observed item responses,  missigness indicators, and individual covariates. The resulting model is a  multidimensional Latent Class IRT (LC-IRT) model (Bartolucci, 2007; von Davier, 2008; Bacci, Bartolucci, and Gnaldi, 2014)  allowing for within-item multidimensionality (Adams, Wilson, and Wang, 1997) in which certain items measure more latent traits.
Here we propose an extension of this model accounting for ordinal item responses and structural missingness (in addition to potentially non-ignorable missingness).

Considering an individual randomly drawn from the population of interest, let $Y_j$ be the ordinal response to item $j=1,\ldots,m$, where the response categories are denoted by integer values $l=1,\ldots, L$. A missing response is denoted with $Y_j={\rm NA}$, which indicates two types of missing:  (i) item $j$ is not due by design (structural missing, thus ignorable); (ii) item $j$ is due but it is skipped (potentially non-ignorable missing). The two types of missing data are distinguished by the response indicator $R_j$, assuming value ${\rm NA}$ if item $j$ is not due,  value $0$ if item $j$ is skipped,  and value $1$ if item $j$ is answered.
Note that the total number of items is $2m$, namely $m$ test items $Y_j$ plus $m$ response indicators $R_j$.

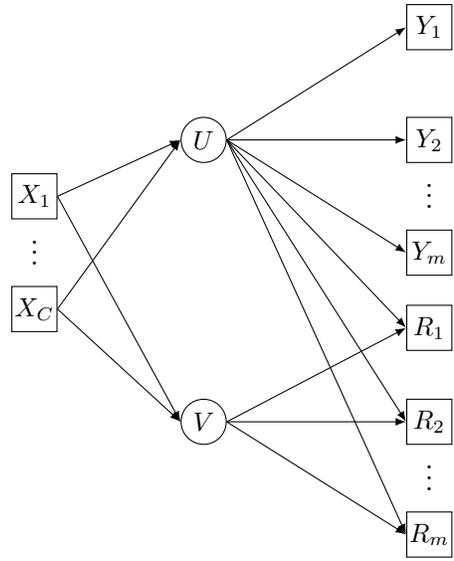
\begin{figure}[!ht]
\begin{center}
\begin{tikzpicture}[scale=1.5]
\node[latent] (U) at (-2,1)  {\footnotesize{$U$}};
\node[manifest] (Y1) at (0,2)  {\footnotesize{$Y_1$}};
\node[manifest] (Y2) at (0,1)  {\footnotesize{$Y_2$}};
\node[manifest] (Ym) at (0,0) {\footnotesize{$Y_m$}};
\node [above] at (0, 0.3) {$\vdots$};

\node[latent] (V) at (-2,-1.5) [shape=circle, draw] {\footnotesize{$V$}};
\node[manifest] (R1) [below of=Ym]  {\footnotesize{$R_1$}};
\node[manifest] (R2) at (0,-1.5) {\footnotesize{$R_2$}};
\node[manifest] (Rm) at (0,-2.5)  {\footnotesize{$R_m$}};
\node [above] at (0, -2.2) {$\vdots$};
\node[manifest] (X1) at (-3.5,0.5)  {\footnotesize{$X_1$}};
\node[manifest] (XC) at(-3.5,-0.5)  {\footnotesize{$X_C$}};
\node [above] at (-3.5,-0.2) {$\vdots$};
\draw [arr]  (U.east) -- (Y2.west);
\draw [arr]  (U.east) -- (Y1.west);
\draw [arr]  (U.east) -- (Ym.west);
\draw [arr]  (U.east) -- (R1.west);
\draw [arr]  (U.east) -- (R2.west);
\draw [arr]  (U.east) -- (Rm.west);
\draw [arr]  (V.east) -- (R1.west);
\draw [arr]  (V.east) -- (R2.west);
\draw [arr]  (V.east) -- (Rm.west);
\draw [arr]  (X1.east) -- (U.west);
\draw [arr]  (X1.east) -- (V.west);
\draw [arr]  (XC.east) -- (U.west);
\draw [arr]  (XC.east) -- (V.west);
\end{tikzpicture}
\caption{\em Path diagram of the LC-IRT model with unidimensional latent traits.}
\label{fig:path}
\end{center}
\end{figure}

The test items $Y_j$, along with the response indicators $R_j$, contribute to measure two latent traits, assumed to be independent given a set of exogenous individual covariates denoted by $\b X = (X_1,\ldots,X_C)\tr$. The first latent trait is described by a multidimensional latent variable $\b U= (U_1, \ldots, U_S)\tr$, representing the abilities measured by the test items $Y_j$. The second latent trait, described by a multidimensional latent variable  $\b V = (V_1, \ldots, V_T)\tr$, represents individual preferences in choosing the test items to answer (i.e., the exam to take) or to skip.
This structure is represented in the path diagram of Figure \ref{fig:path}, which refers to the special case of our application (Section \ref{sec:applic}), where both latent traits have a single component ($S=T=1$).

In the following, we assume that $\b U$ and $\b V$ have a discrete distribution with $k_U$ vectors of support points $\b u_{h_U}=(u_{1h_U}, \ldots, u_{Sh_U})\tr$, $h_U=1,\ldots,k_U$, and $k_V$ vectors of support points $\b v_{h_V}=(v_{1h_V}, \ldots, v_{Th_V})\tr$, $h_V=1,\ldots,k_V$, respectively. This specification corresponds to clustering the individuals into latent classes that are homogeneous with respect to the latent traits.  In the spirit of concomitant variable LC models (Dayton, 1988; Formann, 2007), 
we  allow the membership probabilities of the latent classes to depend on observed covariates through a multinomial logit model (see also Bacci and Bartolucci, 2015):  
\begin{eqnarray}\label{eq:cov1}
\log{\frac{\la_{h_U}(\b x)}{\la_1(\b x)}} &= &
 \b x\tr\b\phi_{h_U},  \quad h_U = 2, \ldots, k_U, \\ 
\log{\frac{\pi_{h_V}(\b x)}{\pi_1(\b x)}} &= &
 \b x\tr\b\psi_{h_V},  \quad h_V = 2, \ldots, k_V,\label{eq:cov2}
\end{eqnarray} 
with $\la_{h_U}(\b x) = Pr(\b U = \b u_{h_U}| \b X = \b x)$ and $\pi_{h_V}(\b x) = Pr(\b V = \b v_{h_V}| \b X = \b x)$,  where the covariate vector $\b x$ includes a constant term, that is, $\b x=(1,x:1,\ldots,x_C)\tr$. The vectors of coefficients $\b\phi_{h_U} = (\phi_{h_U1}, \ldots, \phi_{h_UC})\tr$ and $\b\psi_{h_V} = (\psi_{h_V1}, \ldots, \psi_{h_VC})\tr $ represent the effects of the covariates  on the reference category logits. 

The relationships among the latent variables in $\b U$ and $\b V$ and the item responses $Y_1, \ldots, Y_m$ and the response indicators $R_1, \ldots, R_m$ are described by the measurement part of the model,  specified as a multidimensional LC-IRT model (for details, see Bacci, Bartolucci, and Gnaldi, 2014).  
The proposed model is an extension of the model of Bacci, Bartolucci, and Gnaldi (2014), in that the  multidimensional model structure is  completely general, as it  allows  each indicator to measure one component in $\b U$ and one component in $\b V$ (within-item multidimensionality). As already noted, we extend the model of Bacci and Bartolucci (2015) to ordinal items and structural missing values. 

Let $q_{h_Uh_V,j}=Pr(R_{j} = 1|\b U = \b u_{h_U},\b V= \b v_{h_V})$ denote the probability of answering item $j$ conditionally on $\b U$ and $\b V$. Moreover, let $p_{h_U,jy}=  Pr(Y_j\geq y| \b U = \b u_{h_U})$ denote the probability that the answer to item $Y_j$ is $y$ or higher ($y=2,\ldots,L$), conditionally on the latent trait $\b U$.  In order to select the components of $\b U$ and $\b V$ entering the probabilities  $q_{h_Uh_V,j}$ and $p_{h_U,jy}$, we introduce two sets of indicators $z_{Usj}$ and $z_{Vtj}$, equal to 1 if  item $j$ measures the components $U_s$ and $V_t$, respectively.
Then, we specify a multidimensional LC two-parameter logistic (2PL) model (Bartolucci, 2007) for the probability of answering item $j$:
\begin{equation}
\log\frac{q_{h_Uh_V,j}}{1-q_{h_Uh_V,j}}=  
\gamma_{Uj}\sum_{s=1}^S z_{Usj}u_{sh_U} + \gamma_{Vj}\sum_{t=1}^T z_{Vtj}v_{th_V} - \delta_j, \label{eq:par2}
\end{equation}
where $\delta_j$ can be interpreted as the difficulty to answer item $j$, as higher  values of $\delta_j$ reduces the probability to answer the item; moreover, $\gamma_{Uj}$ and $\gamma_{Vj}$ are discrimination parameters, measuring the effects of the latent traits $\b U$ and $\b V$, respectively, on the probability to answer the item.
Note that, when $\gamma_{Uj} = 0$ for all items, the probabilities of answering the items do not depend on the latent variables in $\b U$, thus the missingness process is ignorable (see the ignorability test of Section \ref{sec:test-ignor}). 
Moreover, the ordinal item responses $Y_j$ are modeled through a graded response parameterization (Samejima, 1969):
\begin{equation}
\log\frac{p_{h_U,jy}}{1-p_{h_U,jy}} = 
\alpha_j \sum_{s=1}^S z_{Usj}u_{sh_U} - \beta_{jy}, \quad  y=2,\ldots,L,
\label{eq:par1}
\end{equation}
where $\beta_{jy}$ is  specific of item $j$ and response category $y$ and it may be interpreted as a difficulty parameter, since higher values of $\beta_{jy}$ ($y=2,\ldots,L$) push the probability distribution of the item  towards the bottom of the scale. On the other end,   $\alpha_j$ is a discrimination parameter, measuring the effect of variables in $\b U$ on the probability distribution of the item. 

In order to ensure the identifiability of the proposed  within-item multidimensional model,  two necessary conditions must be satisfied. The first one requires that at least one item must load only on one of the components of  $\b U$  or only on one of the components of  $\b V$. 
In our specific context related to the treatment of non-ignorable missingness, this condition is always satisfied, as items denoting  responses $Y_1$, \ldots, $Y_m$ measure only latent vector $\b U$. 
Second, suitable  constraints on the item parameters are required. In particular, we constrain one of the discrimination parameters ($\gamma_{Uj}$,  $\gamma_{Vj}$, $\alpha_j$) to be equal to 1 and one difficulty parameter ($\delta_{j}$,  $\beta_{jy}$) to be equal to 0   for each component of each latent variable. Generally speaking, any item  may be  chosen  to be constrained, paying attention to select a different item for each  dimension. In equation \eqref{eq:par2} we constrain $\gamma_{Vj_{t}}=1$ and $\delta_{j_t}=0$, whereas in equation \eqref{eq:par1} we constrain $\alpha_{j_{s}}=1$ and $\beta_{j_{s}1}=0$, with $j_{s}$  and $j_{t}$ ($j_s, j_t=1,\ldots,m$) denoting a specific item $j$, say the first one, which measure components $s$ ($s=1,\ldots,S$) and $t$ ($t=1,\ldots,T$) of $\b U$ and $\b V$, respectively.

The proposed model can be used to predict probabilities for the item result $Y_j$ and response indicator $R_j$ conditionally on specific values of the latent traits $\b U$ and $\b V$. For instance, from equation (\ref{eq:par1}) the predicted probability that item $j$ results in category $y$ is:\begin{align*}
& Pr(Y_j=y| \b U = \b u_{h_U})  =  & \\
 &       = Pr(Y_j\geq y+1| \b U = \b u_{h_U})  - Pr(Y_j\geq y| \b U = \b u_{h_U}) & \\
 &       = \frac{1}{1+\exp[-({\alpha}_j \sum_{s=1}^S z_{Usj}{u}_{sh_U} - {\beta}_{j,y+1})]} - \frac{1}{1+\exp[-({\alpha}_j \sum_{s=1}^S z_{Vsj}{u}_{sh_U} - {\beta}_{jy})]}.
\end{align*}

As another example, from equation (\ref{eq:par2}) the predicted probability that item $j$  is answered turns out to be: 
\begin{align*}
& Pr(R_j=1| \b U = \b u_{h_U},\b V=\b v_{h_V})  =   \\
 &       = \frac{1}{1+\exp[-({\gamma}_{Uj}\sum_{s=1}^S z_{Usj}{u}_{sh_U} + {\gamma}_{Vj}\sum_{t=1}^T z_{Vtj}{v}_{th_V} - {\delta}_j)]}.
\end{align*}
\section{Likelihood inference}\label{sec:inference}
The proposed LC-IRT model under within-item multidimensionality can be estimated through the maximization of the discrete marginal log-likelihood 
\begin{equation}
\label{eq:like}
\ell(\b\eta) = \sum_{i=1}^n \log L_i({\bm y}_{i,obs},\bm r_i|\bm x_i),
\end{equation}
where $\b\eta$ is the vector of model parameters of equations (\ref{eq:cov1}) to (\ref{eq:par1}) further to the support points of $\b U$ and $\b V$,  $\b y_{i,obs}=(y_{i1}, \ldots, y_{im}) \tr$ is the  vector of observed item responses for subject $i$, $\b r_i=(r_{i1}, \ldots, r_{im}) \tr$ is the vector of response indicators for subject $i$, and $\b x_i$ is the vector of covariates for subject $i$. The joint marginal likelihood $L_i({\bm y}_{i,obs},\bm r_i|\bm x_i)$ of subject $i$ in equation (\ref{eq:like}) is given by: 
\begin{equation*}
L_i({\bm y}_{i,obs},\bm r_i|\bm x_i) =  \sum_{h_U=1}^{k_U}\sum_{h_V=1}^{k_V} 
\lambda_{h_U}(\bm x_i)\pi_{h_V}(\bm x_i)p_{h_Uh_V}({\bm y}_{i,obs},\bm r_i),
\end{equation*}
where, given the local independence assumption,
\begin{equation}
\label{eq:jointyr}
 p_{h_Uh_V}({\bm y}_{i,obs},\bm r_i )=\prod_{j=1\,(r_j=1)}^mp_{h_U,jy}\prod_{j=1\,(r_j\neq {\rm NA})}^m q_{h_Uh_V,j}^{r_j}(1-q_{h_Uh_V,j})^{1-r_j}.   
\end{equation} 
Note that if item $j$ is not due, that is, it is missing by design ($r_j={\rm NA}$), it does not contribute to  equation (\ref{eq:jointyr}); while if item $j$ is due but it is skipped ($r_j=0$) it contributes to equation (\ref{eq:jointyr}) only through the term $(1-q_{h_Uh_V,j})$. 

The estimation of the proposed model can be performed by the specific  {\tt R} package  {\tt MLCIRTwithin} (Bartolucci and Bacci, 2015), which
 maximizes the marginal likelihood (\ref{eq:like}) through the EM algorithm (Dempster, Laird, and Rubin, 1977), following the same lines as in Bacci and Bartolucci (2015). Moreover, it allows for several options, such as:  
different number of latent classes for the two latent variables, binary or ordinal items for both the item response process and the missingness process,
  Rasch or 2PL parameterization for binary items, graded response or partial credit (Masters, 1982) parameterization  for ordinal items,  
 multinomial logit or global logit parameterization (Agresti, 2002) for the sub-model that explains the effect of covariates on the probabilities. 
Under the assumption of normally distributed latent traits, the estimation of within-item multidimensional IRT models assuming the presence of a general latent trait affecting all items, according to the formulation of Gibbons and Hedeker (1992), Gibbons et al. (2007), and Cai (2010), can be performed by means of the {\tt R} package {\tt mirt} (Chalmers, 2012). This package also admits discrete latent variables, but in this case it has a  limited flexibility, as the same number of latent classes is required for the two latent traits (i.e., $k_U=k_V$).

For model selection we rely on information criteria, such as the Bayesian Information Criterion (BIC; Schwarz, 1978), to compare non-nested models (mainly, for the choice of the number of support points $k_U$ and $k_V$), while we use the likelihood ratio test to compare nested models. 

In order to facilitate the interpretation of the results and the comparison of models with different specifications, we standardize the support points as follows
     \begin{eqnarray}
     \label{eq:stdpoints} 
    \hat{u}^*_{sh_U} & = & \frac{\hat{u}_{sh_U}-\hat{\mu}_{U_s}}{\hat{\sigma}_{U_s}}, \quad s=1,\ldots, S,\\ \nonumber
    \hat{v}^*_{th_V} & = & \frac{\hat{v}_{th_V}-\hat{\mu}_{V_t}}{\hat{\sigma}_{V_t}},  \quad t=1,\ldots, T, \nonumber  
    \end{eqnarray} 
where $\hat{\mu}_{U_s}$ and $\hat{\sigma}_{U_s}$ are the mean and the standard deviation of $\hat{u}_{s1}, \ldots, \hat{u}_{sk_U}$, whereas  $\hat{\mu}_{V_t}$ and $\hat{\sigma}_{V_t}$ are the mean and the standard deviation of $\hat{v}_{t1}, \ldots, \hat{v}_{tk_V}$. The item parameters of equations \eqref{eq:par2} and \eqref{eq:par1} must be transformed coherently as follows:
     \begin{eqnarray}
     \label{eq:stdpar} \nonumber
   \hat{\alpha}^*_{j} & = &  \hat{\alpha}_{j}\sum_{s=1}^S z_{Usj}\hat{\sigma}_{U_s}, \\ \nonumber
   \hat{\beta}^*_{jy} & = &  \hat{\beta}_{jy} - \hat{\alpha}_j \sum_{s=1}^S z_{Usj}\hat{\mu}_{U_s}, \\ 
     \hat{\gamma}^*_{1j} & = &  \hat{\gamma}_{1j}\sum_{s=1}^S z_{Usj}\hat{\sigma}_{U_s}, \\ \nonumber
    \hat{\gamma}^*_{2j}  & = &  \hat{\gamma}_{2j}\sum_{t=1}^T z_{Vtj}\hat{\sigma}_{V_t},  \\ \nonumber
    \hat{\delta}^*_{j}  & = &  \hat{\delta}_{j}  - \hat{\gamma}_{1j}\sum_{s=1}^S z_{Usj}\hat{\mu}_{U_s} - \hat{\gamma}_{2j}\sum_{t=1}^T z_{Vtj}\hat{\mu}_{V_t}. \nonumber
    \end{eqnarray} 

The standard errors of the transformed item parameters are obtained through the Delta method (Casella and Berger, 2006).

\section{Analysis of student careers}\label{sec:applic}\vspace*{-0.25cm}
We applied the LC-IRT model described in Section \ref{sec:mod} to the analysis of the performance of university students described in Section \ref{sec:data}. In the following we illustrate model specification and fitting, performing several tests for model selection, including a test for the  ignorability of the missing data mechanism. We report estimates of model parameters, or suitable transformations improving interpretations. We  discuss the main results, focusing on the discrimination and difficulty of the exams and   the interpretation of the latent structure.

\subsection{Model specification}
For the analysis of the performance of university students we assumed $S=1$, that is, all exams measure the same latent ability ($U_s=U$), and $T=1$, that is, there is one latent tendency to take an exam ($V_t=V$). Besides, we took explicitly into account that, for each of the six exams, there are four teachers and each of them  defines a group, whose assignment to students depends on the first letter of the surname. Consequently, each combination exam-by-group defines a different item and the total number of items is therefore $m=24$. 

The tendency to take an exam, corresponding to $V$, is measured by the binary variable $R_j$ for $j=1,\ldots,24$ that is observed for a given student when item $j$ corresponds to the group to which he/she belongs to; otherwise $R_j$ is missing by design. Given that $R_j$ is observed, it equals $1$ if the student enrolls to the corresponding exam at least once during the year and it equals 0 if the student skips the exam.
Skipping the exam may depend both on the tendency $V$ to take an exam and on the ability $U$ that university exams  contribute to measure. The structure of the proposed model is illustrated by the path diagram of Figure \ref{fig:path}. 

Conditional to the enrollment, the student can fail  or pass the exam with a given grade, ranging from 18 to 30, plus 30 with honors. We outline that 
 the  distribution of exam grades is far from to be normal, with peaks of observations in correspondence to certain grades (Bertaccini, Grilli, and Rampichini, 2013) and the maximum grade standing out of a quantitative scale. Moreover, grades from 0 to 17, denoting the exam failure, are not observable. 
Thus, the result on exam $j$ is defined by the ordinal variable  $Y_j$ with categories defined as follows:
\begin{eqnarray*}
\left\{
\begin{array}{lcl}
Y_j=NA & \textrm{if} & R_j=0,\\
Y_j = 0 & \textrm{if} & R_j=1\; {\rm and}\;Z_j = {\rm NA},\\
Y_j=1 &  \textrm{if} & 18 \leq Z_j \leq 21,\\
Y_j=2 &  \textrm{if} & 22 \leq Z_j \leq 24,\\
Y_j = 3 & \textrm{if} & 25 \leq Z_j \leq 27, \\
Y_j = 4 & \textrm{if} & Z_j \geq 28, \\
\end{array}
\right.
\end{eqnarray*}
where $Z_j$ is the exam grade in the original scale, with  $Z_j \ge 18$ if the exam is passed, and  $Z_j = {\rm NA}$ if the exam is failed. 

We  specified a multinomial logit model for the effect of the observed student characteristics (i.e., degree program, gender, HS grade, HS type, and late matriculation) on the probabilities of the latent variables $U$ in equation \eqref{eq:cov1} and $V$ in equation \eqref{eq:cov2}.
Moreover, we specified a graded response model as in equation \eqref{eq:par1} for the exam result $Y_j$, and a 2PL model as in equation \eqref{eq:par2} for the enrollment $R_j$. 

\subsection{Model fitting}\label{sec:BIC}
In order to select the number of latent classes of $U$ and $V$, we fitted a series of models with covariates, making comparisons through the BIC index.
These models are fitted as described in section \ref{sec:inference}. 
As a first step, we considered values of $k_U$ and $k_V$ equal or greater than 2, so as to faithfully reflect the latent structure described by the path diagram in Figure \ref{fig:path}. According to the results reported in Table \ref{tab:bic}, we selected $k_U=4$ latent classes for $U$, and $k_V=2$ latent classes for $V$.

\begin{table}[!ht] 
\begin{center}    
\caption{\em Selection of latent classes.}\label{tab:bic}
\small
\begin{tabular}{cc|ccc}
\hline\hline
$k_U$   &   $k_V$   &   $\hat{\ell}$    &   \# par  &   BIC \\
\hline
2   &   2   &   -6520.37    &   208 & 14446.41    \\
2   &   3   &    -6505.63   &   217 &  14477.76   \\
3   &   2   &   -6387.18    &   217 & 14240.87    \\
3   &   3   &   -6364.32    &   226 & 14255.96    \\
4   &   2   &    -6338.27  &   226 &  14203.86   \\
4   &  3    &   -6325.72  & 235 & 14239.58 \\
5   &  2   &    -6323.84   &    235 & 14235.84 \\
5   &  3    & -6304.16  &  244  & 14257.30 \\
\hline
\end{tabular}   
\end{center}
\end{table}

In order to check for local maxima, we repeated the model estimation process for different random starting values of the parameters.

\subsection{Testing the ignorability of the missing data mechanism}
\label{sec:test-ignor}
In our setting, missing data regarding variable $Y_j$ are generated by the student decision to not take an exam.
The specified model assumes that the choice to take an exam depends both on a latent variable  representing the ``temperament'' of a student (describing his/her propensity to enroll) and on the student's ability. The dependence on the ability amounts to treat the missing data mechanism as non-ignorable (see Section \ref{sec:mod}).

To test the ignorability assumption we compared  the proposed multidimensional LC-IRT model with a restricted model where  exam enrollment does not depend on the ability $U$, namely we tested the hypothesis $\gamma_{Uj}=0, \forall j=1, \ldots, 24$.

The likelihood-ratio test (LRT) statistic is $LRT=2 \times(6533.720-6338.268)=390.904$,  with $24$ degrees of freedom yielding a very low $p$-value.
Therefore we proceeded with the proposed multidimensional LC-IRT   model accounting for the non-ignorable missing mechanism.

\subsection{Discrimination and difficulty of the exams}\label{sec:diff-discr}
Table \ref{tab:discr}  reports the discrimination parameters of equation \eqref{eq:par1} for the exam outcome $Y_j$, and the discrimination parameters of equation \eqref{eq:par2} for exam enrollment $R_j$. In order to increase the interpretability of the results, all the parameters reported in Table \ref{tab:discr} are  scaled according to equations (\ref{eq:stdpar}).

\begin{table}[!ht] 
\begin{center}    
\caption{{\em Estimated scaled discrimination item parameters.}}   
\label{tab:discr}    
\small
\begin{tabular}{ll|rrr|rrr|rrr}
\hline\hline
\multicolumn{2}{c|}{Item}&    \multicolumn3c{ $U \rightarrow Y_j$ } & \multicolumn3c{ $U \rightarrow R_j$ } & \multicolumn3c{ $V \rightarrow R_j$ } \\ \hline
Course                & Group           &   $\hat{\alpha}_j^*$  &   $\hat{\textrm{se}}_{{\alpha}_{j}^*}$    &   $p$-value   &   $\hat{\gamma}_{Uj}^*$   &   $\hat{\textrm{se}}_{{\gamma}_{Uj}^*}$   &   $p$-value   &   $\hat{\gamma}_{Vj}^*$   &   $\hat{\textrm{se}}_{{\gamma}_{Vj}^*}$   &   $p$-value   \\
    \hline
Accounting &  A-C   &   2.127   &   0.285   &   $<$0.001&   0.904   &   0.375   &   0.016   &   0.401   &   0.274   &   0.144   \\
 &  D-L &   1.795   &   0.259   &   $<$0.001    &   0.535   &   0.349   &   0.125   &   0.594   &   0.433   &   0.170   \\
 &  M-P &   2.527   &   0.380   &   $<$0.001    &   1.172   &   0.533   &   0.028   &   -0.503  &   0.525   &   0.338   \\
 &  Q-Z &   2.337   &   0.357   &   $<$0.001    &   0.433   &   0.318   &   0.173   &   0.522   &   0.311   &   0.093   \\
    \hline
Mathematics &  A-C  &   2.241   &   0.410   &   $<$0.001    &   1.918   &   0.893   &   0.032   &   2.448   &   1.176   &   0.037   \\
 &  D-L &   2.134   &   0.386   &   $<$0.001    &   1.278   &   0.440   &   0.004   &   2.251   &   0.772   &   0.004   \\
 &  M-P &   1.731   &   0.402   &   $<$0.001    &   1.700   &   0.503   &   0.001   &   1.782   &   0.587   &   0.002   \\
 &  Q-Z &   2.963   &   0.707   &   $<$0.001    &   5.050   &   4.605   &   0.273   &   6.783   &   5.266   &   0.198   \\
    \hline
Law &  A-C  &   1.849   &   0.427   &   $<$0.001    &   0.903   &   0.196   &   $<$0.001    &   -0.380  &   0.221   &   0.085   \\
 &  D-L &   3.016   &   0.506   &   $<$0.001    &   1.303   &   0.254   &   $<$0.001    &   -0.390  &   0.267   &   0.144   \\
 &  M-P &   1.391   &   0.306   &   $<$0.001    &   1.144   &   0.267   &   $<$0.001    &   -0.804  &   0.314   &   0.011   \\
 &  Q-Z &   1.783   &   0.425   &   $<$0.001    &   1.033   &   0.241   &   $<$0.001    &   -0.279  &   0.214   &   0.192   \\
    \hline
Management  &  Busi A-L  &   2.990   &   0.530   &   $<$0.001    &   2.287   &   0.389   &   $<$0.001    &   0.675   &   0.315   &   0.032   \\
   & Busi  M-Z   &   3.163   &   0.484   &   $<$0.001    &   1.339   &   0.249   &   $<$0.001    &   -0.304  &   0.221   &   0.169   \\
  & Econ A-L   &   1.976   &   0.389   &   $<$0.001    &   1.106   &   0.250   &   $<$0.001    &   0.539   &   0.296   &   0.068   \\
  & Econ M-Z   &   0.662   &   0.306   &   0.030   &   2.212   &   0.500   &   $<$0.001    &   0.605   &   0.455   &   0.184   \\
    \hline
MicroEcon &  A-C    &   1.249   &   0.325   &   $<$0.001    &   1.429   &   0.247   &   $<$0.001    &   0.310   &   0.238   &   0.193   \\
 &  D-L &   1.130   &   0.402   &   0.005   &   3.114   &   0.598   &   $<$0.001    &   0.450   &   0.319   &   0.159   \\
 &  M-P &   1.822   &   0.366   &   $<$0.001    &   1.889   &   0.353   &   $<$0.001    &   -0.447  &   0.294   &   0.128   \\
 &  Q-Z &   2.350   &   0.588   &   $<$0.001    &   2.202   &   0.440   &   $<$0.001    &   0.926   &   0.282   &   0.001   \\
    \hline
Statistics &  A-C   &   2.787   &   0.445   &   $<$0.001    &   2.333   &   0.466   &   $<$0.001    &   0.946   &   0.387   &   0.014   \\
 &  D-L &   2.496   &   0.389   &   $<$0.001    &   1.567   &   0.290   &   $<$0.001    &   0.808   &   0.295   &   0.006   \\
 &  M-P &   2.867   &   0.497   &   $<$0.001    &   1.772   &   0.319   &   $<$0.001    &   0.104   &   0.292   &   0.722   \\
 &  Q-Z &   2.258   &   0.468   &   $<$0.001    &   1.998   &   0.469   &   $<$0.001    &   1.020   &   0.347   &   0.003   \\
         \hline
\end{tabular}   
\end{center}
\end{table}

Note that all the discrimination parameters $\hat{\alpha}_j^*$ relating exam results $Y_j$ to the ability $U$ are significantly different from zero, namely all the exams contribute to measure the latent ability.
Accounting, Mathematics, and Statistics tend to have a higher discrimination power, that is, the results of these exams are more sensitive to variations in student ability. However, there are differences across groups of the same course, especially for Law and Management. 

According to equation (\ref{eq:par2}), enrollment to an exam $R_j$ is affected by the student's ability $U$ through the $\gamma_{Uj}^*$ parameters, and by the latent variable $V$ through the $\gamma_{Vj}^*$ parameters.
The effect of the latent variable $V$ is positive for Mathematics and Statistics, and negative for Law; thus $V$ can be interpreted as the tendency of the student to take exams in quantitative subjects as opposed to exams in qualitative subjects.
Considering statistical significance at $5\%$, student's ability $U$ significantly affects the enrollment for most exams, whereas student's tendency $V$ has a significant effect for just around one-third of the items.
Moreover, comparing the absolute value of $\hat{\gamma}_{1j}^*$ and $\hat{\gamma}_{2j}^*$ it turns out that the enrollment to exam is affected more by $U$ than by $V$, with the notable exception of Mathematics. 

The dependence of the $R_j$ variables for the exam enrollment on the ability $U$ suggests that a model for evaluating the student's proficiency should account for enrollment decisions; in statistical terms, this provides evidence that the enrollment process generating missing exam grades is not ignorable, as confirmed by the likelihood-ratio test reported  in Section \ref{sec:test-ignor}. 

Equations \eqref{eq:par2} and \eqref{eq:par1} include five difficulty parameters for each item $j$, specifically  four parameters $\beta_{1j}^* \ldots \beta_{4j}^*$ for each exam result $Y_j$, and parameter $\delta_j^*$ for each exam enrollment variable $R_j$. The estimates of the difficulty parameters (shown in the online Supplementary Material, Tables 1-2) are not easily interpretable, thus we converted such parameters into probabilities.
In particular, Table  \ref{tab:proby} reports the probabilities of exam results for a student with average ability, that is, $U=0$. In addition, the right part of  Table \ref{tab:proby} reports the conditional probability of passing the exam $Pr(Y_j > 0|U)$ for certain  values of the student's ability, that is, $U= -\sigma_U$, $U=0$, and $U= +\sigma_U$, with $\sigma_U$ denoting the estimated standard deviation of student ability $U$.

\begin{table}[!ht] 
\begin{center}    
\caption{\em Predicted probabilities of exam result $Y_j$ at some values of latent ability $U$.}
\label{tab:proby}       
\small
\begin{tabular}{ll|rrrrr|rrrr}
\hline\hline
\multicolumn{2}{c|}{Item}&    \multicolumn{5}{c|}{ {$Pr(Y_j=y \mid U=0)$}}&\multicolumn{4}{c}{Passing rate $Pr(Y_j>0\mid U)$}\\ \cline{3-11}
Course                & Group                            &\multicolumn{1}{c}{0}&\multicolumn{1}{c}{1}&\multicolumn{1}{c}{2}&\multicolumn{1}{c}{3}&\multicolumn{1}{c|}{4}&\multicolumn{3}{c}{$U=u$}\\ \cline{8-10}
                      &                                  &Failed&$18$-$21$&$22$-$24$&$25$-$27$&$\geq 28$&$-\sigma_U$&\multicolumn{1}{c}{0}&$+\sigma_U$&range\\ \hline
Accounting  &   A-C &   0.348   &   0.328   &   0.172   &   0.124   &   0.028   &   0.183   &   0.652   &   0.940   &   0.758   \\
    &   D-L &   0.648   &   0.168   &   0.126   &   0.052   &   0.006   &   0.083   &   0.352   &   0.766   &   0.683   \\
    &   M-P &   0.378   &   0.219   &   0.305   &   0.094   &   0.005   &   0.116   &   0.622   &   0.954   &   0.837   \\
    &   Q-Z &   0.141   &   0.295   &   0.355   &   0.176   &   0.033   &   0.371   &   0.859   &   0.984   &   0.614   \\
\hline
Mathematics &   A-C &   0.839   &   0.114   &   0.029   &   0.012   &   0.005   &   0.020   &   0.161   &   0.643   &   0.623   \\
    &   D-L &   0.803   &   0.136   &   0.037   &   0.019   &   0.005   &   0.028   &   0.197   &   0.674   &   0.646   \\
    &   M-P &   0.872   &   0.069   &   0.036   &   0.016   &   0.007   &   0.025   &   0.128   &   0.453   &   0.427   \\
    &   Q-Z &   0.906   &   0.085   &   0.005   &   0.004   &   0.000   &   0.005   &   0.094   &   0.667   &   0.662   \\
\hline
Law &   A-C &   0.824   &   0.105   &   0.060   &   0.008   &   0.004   &   0.033   &   0.176   &   0.576   &   0.544   \\
    &   D-L &   0.530   &   0.168   &   0.237   &   0.057   &   0.008   &   0.042   &   0.470   &   0.948   &   0.906   \\
    &   M-P &   0.709   &   0.142   &   0.042   &   0.092   &   0.016   &   0.093   &   0.291   &   0.623   &   0.530   \\
    &   Q-Z &   0.480   &   0.259   &   0.177   &   0.070   &   0.014   &   0.154   &   0.520   &   0.866   &   0.711   \\
\hline
Management  &   Busi A-L &   0.222   &   0.187   &   0.350   &   0.181   &   0.061   &   0.150   &   0.778   &   0.986   &   0.836   \\
    &   Busi M-Z &   0.761   &   0.063   &   0.131   &   0.037   &   0.008   &   0.013   &   0.239   &   0.881   &   0.868   \\
    &   Econ  A-L  &   0.381   &   0.204   &   0.209   &   0.163   &   0.043   &   0.184   &   0.619   &   0.921   &   0.738   \\
    &   Econ M-Z   &   0.101   &   0.026   &   0.093   &   0.571   &   0.209   &   0.821   &   0.899   &   0.945   &   0.124   \\
\hline
MicroEcon   &   A-C &   0.707   &   0.124   &   0.102   &   0.049   &   0.018   &   0.106   &   0.293   &   0.591   &   0.485   \\
    &   D-L &   0.809   &   0.046   &   0.046   &   0.063   &   0.036   &   0.071   &   0.191   &   0.422   &   0.351   \\
    &   M-P &   0.355   &   0.157   &   0.161   &   0.248   &   0.078   &   0.227   &   0.645   &   0.918   &   0.691   \\
    &   Q-Z &   0.650   &   0.117   &   0.067   &   0.081   &   0.085   &   0.049   &   0.350   &   0.850   &   0.801   \\
\hline
Statistics  &   A-C &   0.501   &   0.292   &   0.120   &   0.055   &   0.032   &   0.058   &   0.499   &   0.942   &   0.884   \\
    &   D-L &   0.547   &   0.231   &   0.157   &   0.041   &   0.024   &   0.064   &   0.453   &   0.909   &   0.846   \\
    &   M-P &   0.673   &   0.197   &   0.076   &   0.038   &   0.017   &   0.027   &   0.327   &   0.895   &   0.868   \\
    &   Q-Z &   0.625   &   0.197   &   0.057   &   0.073   &   0.048   &   0.059   &   0.375   &   0.852   &   0.793   \\
\hline
\end{tabular}   
\end{center}
\end{table}

We note large variability among courses and, in some cases, also across groups of the same course. For the majority of items, the most likely result is a failure. 
Moreover, for the majority of the courses the modal grade of  passed exams is $18-21$, with some notable exceptions, such as Management Econ M-Z. 
The values of the discrimination parameters $\hat{\alpha}_j^*$ imply that the probability to pass the exam depends on the student's ability: the range reported in the last column of Table \ref{tab:proby} is large, with relevant differences both within and between courses.

The probabilities reported in Table \ref{tab:proby} can be used to predict the performance of a student with a hypothetical ability $U$, depending on the chosen degree program  and the group assigned on the basis on the first letter of the surname.  For example, for a student with a high level of ability (say, $+\sigma_U$), enrolled in the degree program \emph{Business} and belonging to the A-C group, the  probability to pass all the exams can be obtained by multiplying the six probabilities $Pr(Y_j>0\mid U=+\sigma_U)$ corresponding to the A-C group, that is,  $0.940 \times 0.643 \times \ldots \times 0.942 =0.191$. It is worth noting that the same probability rises to 0.362 for a student belonging to the Q-Z group, whereas it drops to 0.173 for a student enrolled in the degree program \emph{Economics} and belonging to the D-L group. Similar computations show that the probability to pass all six exams is less that $0.01$ for students with average ability ($U=0$).

In a similar way we can compute the probability of other patterns. For example, the probability to pass only   Accounting for a student with ability level equal to the average, who is enrolled in the degree program \emph{Business} and belongs to the D-L group, is $0.352 \times (1-0.197) \times \ldots \times (1-0.453) =0.015$;  it rises to 0.025 for a colleague belonging to the same group but enrolled in the degree program \emph{Economics} and to 0.070 for a colleague enrolled in the same degree program but belonging to group M-P.  
Moreover, for a student with a low ability level (say, $-\sigma_U$), enrolled in the degree program \emph{Business} and belonging to the D-L group, the probability to pass one exam out of six is $0.306$, obtained by adding the probability to pass only Accounting, the probability to pass only Mathematics, and so on. For a similar student belonging to group M-P the probability to pass only one exam rises to 0.349.

Table \ref{tab:probr} reports the probability to enroll in an exam for some values of latent ability $U$ and latent tendency $V$.
The first column of  Table \ref{tab:probr} reports the probabilities for a student with average values for both latent variables ($U=0, V=0$), thus depending only on the estimated difficulty parameters $\hat{\delta}_j^*$.

\begin{table}[!ht] 
\begin{center}    
\caption{\em Predicted probabilities of enrollment $R_j$ at some values of latent ability $U$ and latent tendency $V$.}     
\label{tab:probr}  
\small  
\begin{tabular}{ll|rrrrrrrr}
\hline\hline
\multicolumn{2}{c|}{Item}&  &  \multicolumn{5}{c}{ {$Pr(R_j=1 \mid U=u, V=v)$}} &&\\ \cline{3-10}
Course                & Class&$u$&\multicolumn{1}{|c}{0}&$-\sigma_U$&$+\sigma_U$&\multicolumn{1}{c}{0}&\multicolumn{1}{c|}{0} &\multicolumn{2}{c}{Range}\\ \cline{3-10}
                      &      &$v$&\multicolumn{1}{|c}{0}&\multicolumn{1}{c}{0}          &\multicolumn{1}{c}{0}          &$-\sigma_V$&\multicolumn{1}{c|}{$+\sigma_V$}& $\pm\sigma_U$&$\pm\sigma_V$\\ \hline
Accounting  &   A-C &&0.962 &   0.910   &   0.984   &   0.944   &   0.974   &   0.074   &   0.030   \\
    &   D-L &   &0.952  &   0.921   &   0.971   &   0.916   &   0.973   &   0.050   &   0.056   \\
    &   M-P &   &0.973  &   0.919   &   0.992   &   0.984   &   0.957   &   0.073   &   -0.027  \\
    &   Q-Z &   &0.913  &   0.872   &   0.942   &   0.861   &   0.946   &   0.070   &   0.085   \\
\hline
Mathematics &   A-C &&  0.797   &   0.367   &   0.964   &   0.254   &   0.979   &   0.597   &   0.724   \\
    &   D-L &   &0.785  &   0.505   &   0.929   &   0.278   &   0.972   &   0.425   &   0.694   \\
    &   M-P &&  0.768   &   0.376   &   0.948   &   0.357   &   0.952   &   0.571   &   0.594   \\
    &   Q-Z &&  0.905   &   0.058   &   0.999   &   0.011   &   1.000   &   0.942   &   0.989   \\
\hline
Law &   A-C &&  0.323   &   0.162   &   0.541   &   0.411   &   0.246   &   0.379   &   -0.165  \\
    &   D-L &&  0.506   &   0.218   &   0.790   &   0.602   &   0.409   &   0.572   &   -0.192  \\
    &   M-P &&  0.580   &   0.306   &   0.813   &   0.755   &   0.382   &   0.507   &   -0.373  \\
    &   Q-Z &&  0.564   &   0.315   &   0.784   &   0.631   &   0.495   &   0.469   &   -0.137  \\
\hline
Management  &   Busi A-L &&  0.879   &   0.424   &   0.986   &   0.787   &   0.934   &   0.562   &   0.148   \\
    &   Busi M-Z &&  0.792   &   0.499   &   0.936   &   0.837   &   0.737   &   0.437   &   -0.100  \\
    &   Econ  A-L  &&  0.644   &   0.374   &   0.845   &   0.513   &   0.756   &   0.471   &   0.243   \\
    &   Econ M-Z   &&  0.885   &   0.456   &   0.986   &   0.807   &   0.933   &   0.530   &   0.126   \\
\hline
MicroEcon   &A-C    &&  0.268   &   0.081   &   0.605   &   0.212   &   0.333   &   0.524   &   0.121   \\
    &   D-L &&  0.144   &   0.007   &   0.791   &   0.097   &   0.209   &   0.784   &   0.112   \\
    &   M-P &&  0.599   &   0.184   &   0.908   &   0.700   &   0.488   &   0.724   &   -0.212  \\
    &   Q-Z &&  0.526   &   0.109   &   0.909   &   0.305   &   0.737   &   0.800   &   0.431   \\
\hline
Statistics  &   A-C &&  0.852   &   0.358   &   0.983   &   0.691   &   0.937   &   0.625   &   0.246   \\
    &   D-L &&  0.679   &   0.306   &   0.910   &   0.485   &   0.826   &   0.604   &   0.341   \\
    &   M-P &&  0.690   &   0.275   &   0.929   &   0.667   &   0.712   &   0.654   &   0.044   \\
    &   Q-Z &&  0.866   &   0.466   &   0.979   &   0.699   &   0.947   &   0.513   &   0.248   \\
\hline
\end{tabular}   
\end{center}
\end{table}

Similarly to the exam result $Y_j$, the enrollment in the exam $R_j$ shows a large variability among courses and, in some cases, also across groups of the same course. The  enrollment rate is high for Accounting and low for Microeconomics and Law. Moreover,  Microeconomics shows large differences  between groups, ranging from 0.14 to 0.60.
Looking at the last two columns of Table \ref{tab:probr}, we see that the probability to enroll in the exam depends more on student ability $U$ than on tendency $V$, with the exception of Mathematics. The effect of $V$ is relevant and positive for Mathematics and Statistics and negative for Law, thus confirming the interpretation of $V$ in terms of tendency to take quantitative exams.

\subsection{Estimated latent structure and covariate effects}

Table \ref{tab:support} reports the estimated support points and corresponding average probabilities for the latent classes of ability $U$ and  tendency $V$. The support points $\hat{u}^*_{h_U}$ and $\hat{v}^*_{h_V}$ are standardized as described by equations (\ref{eq:stdpoints}).

\begin{table}[!ht] 
\begin{center}    
\caption{\em Estimated support points and corresponding average probabilities for the latent classes of ability $U$ and  tendency  $V$.}
\label{tab:support} 
\small
\begin{tabular}{l|rrrr|rr}
\hline\hline
    &        \multicolumn{4}{c|}{Ability $U$} &        \multicolumn{2}{c}{Tendency $V$} \\ 
    &        \multicolumn{4}{c|}{latent class $h_U$} &        \multicolumn{2}{c}{latent class $h_V$} \\ \cline{2-7}
    &   $1$& $2$&$3$&$4$&$1$&$2$  \\ \hline

Support points ($\hat{u}^*_{h_U}$, $\hat{v}^*_{h_V}$)          &-1.485 & -0.129  &0.784   & 1.937  &-0.949 & 1.054 \\
Average probabilities ($\overline{\lambda}_{h_U}$, $\overline{\pi}_{h_V}$) & 0.228& 0.395  &0.294   & 0.083  & 0.526 & 0.474 \\
\hline
 \end{tabular}   
\end{center}
\end{table}

Table \ref{tab:regress} reports the estimated coefficients $\hat{\phi}_{h_Uc}$ ($h_U=2, 3, 4$) of the multinomial logit model (\ref{eq:cov1}) for the probabilities $\lambda_{h_U}$  of the latent ability $U$, and  the estimated coefficients $\hat{\psi}_{h_Vc}$ ($h_V=2$), of the multinomial logit model (\ref{eq:cov2}) for the probabilities  $\pi_{h_V}$ of the latent tendency $V$. Note that, since $V$ has only two latent classes, the multinomial logit model (\ref{eq:cov2})  reduces to a binary logit model.

\begin{table}[!ht] 
\begin{center}    
\caption{\em Estimated coefficients of the multinomial logit models for the probabilities of the latent classes of ability $U$ and  tendency  $V$.}
\label{tab:regress} 
\small
\begin{tabular}{l|ccc|c}
\hline\hline
 & \multicolumn3{c|}{Ability $U$}   &  \multicolumn1c{Tendency $V$}  \\
&  $\hat{\phi}_{2c}$ &  $\hat{\phi}_{3c}$ &  $\hat{\phi}_{4c}$  &      \multicolumn1c{$\hat{\psi}_{2c}$} \\
\hline
Constant    &   \;\;\;0.748$^*$ &   \;0.568 &   \;\;-1.956$^*$  &   \;\;-0.869$^*$  \\
Degree Economics   &   -0.434  &   -0.030  &   \;0.045 &   \;\;\;0.716$^*$ \\
Female  &   \;0.434 &   \;0.059 &   -0.487  &   -0.147  \\
HS grade $\geq 80$  &   \;0.014 &   \;\;\;0.118$^*$ &   \;\;\;0.265$^*$ &   -0.020  \\
HS type (ref:technical)     &       &       &       &       \\
\hspace{2 ex} HS humanities     &   \;0.138 &   \;0.148 &   \;0.691 &   -0.240  \\
\hspace{2 ex} HS scientific &   -0.061  &   \;\;\;1.020$^*$ &   \;\;\;2.219$^*$ &   \;\;\;1.963$^*$ \\
\hspace{2 ex} HS other  &   -0.163  &   -0.163  &   -0.282  &   -0.336  \\
Late matriculation &   -0.191  &   \;\;-1.415$^*$  &   \;\;-1.834$^*$  &   \;\;-0.744$^*$  \\
 \hline
\multicolumn{5}{l}{\small Parameters with $^*$ have $p$-value$<0.05$}.
\end{tabular}   
\end{center}
\end{table}

The distribution of the ability $U$ has four support points and it is right skewed. The average probability of the first class ($\overline{\lambda}_{1}=0.228$) is similar to the observed proportion of students who did not pass any exam (0.243). The last class is the smallest one ($\overline{\lambda}_{4}=0.083$) and it includes very good students, with an ability equal to about two standard deviations above the mean. Some student's characteristics have a significant effect on the ability (Table \ref{tab:regress}): students with a higher grade and students with a scientific HS degree tend to belong to latent classes of higher ability (i.e., classes 3 and 4), while the reverse holds for late matriculated students. 

The distribution of the tendency $V$ gives rise to two latent classes of similar size, with support points -0.949 and 1.054. As noted in Section \ref{sec:diff-discr}, students belonging to the second latent class prefer to take quantitative exams. For a baseline student (degree in {\em  Business}, male, HS grade at a mid-point, HS type technical, no late matriculation), the predicted probability to belong to the second class  is  0.295. 
This probability raises to 0.462 for a baseline student but enrolled in the degree of {\em Economics} and to 0.749 for a baseline student but with a scientific HS degree. On the contrary,  this probability decreases to 0.166 for a late matriculated student.

\subsection{Testing differences across groups of the same course}

The results of Section \ref{sec:diff-discr} show that, for some courses, discrimination and difficulty are markedly different across the four groups.
In the   model described by equations \eqref{eq:par2} and \eqref{eq:par1}, an item $j$ corresponds to a group of a given course, for example $j=1$ corresponds to group A-C of Accounting. The model has eight parameters for each item $j$: three discrimination parameters  $(\alpha_j, \gamma_{Uj}, \gamma_{Vj})$ and five difficulty parameters $(\beta_{1j},\beta_{2j},\beta_{3j},\beta_{4j},\delta_{j})$.

A test of homogeneity for the four groups of a given course can be performed comparing the full model with a restricted model, where the items corresponding to the four groups have the same set of parameters.  For example, for the course of Accounting the restricted model assumes $\alpha_{1}=\alpha_{2}=\alpha_{3}=\alpha_{4}$, and similarly for the other parameters, for a total of $3 \times 8=24$ restrictions.

Table \ref{tab:lr-groups} reports the LRT statistics comparing the full model with a restricted model for each course, collapsing the items corresponding to different groups.
These statistics can be interpreted as an indicator of differences among groups of a given course. 

\begin{table}[!ht]
\begin{center}    
\caption{\em Likelihood-ratio test comparing full model with models collapsing groups.}
\label{tab:lr-groups} 
\small
\begin{tabular}{l|rrrrr}
\hline\hline
Exam        &   $logL$    &   \# par  &   LRT stat.  &   df  &   $p$-value   \\
\hline
Full model  &   -6338.27    &   226 &   --  &   --  &   --  \\
Accounting  &   -6389.59    &   202 &   102.64  &   24  &   0.000   \\
Mathematics &   -6349.62    &   202 &   22.70   &   24  &   0.538   \\
Law &   -6397.06    &   202 &   117.59  &   24  &   0.000   \\
Management  &   -6464.91    &   202 &   253.29  &   24  &   0.000   \\
MicroEcon   &   -6411.62    &   202 &   146.71  &   24  &   0.000   \\
Statistics  &   -6358.22    &   202 &   39.90   &   24  &   0.022   \\
\hline
\end{tabular}   
\end{center}
\end{table}

Mathematics shows the lowest value, followed by Statistics, while Management has the highest value. All test statistics deal to reject the homogeneity assumption, except for Mathematics, thus confirming the appropriateness of  a model treating the groups as distinct items with certain structural missing values.

\subsection{Sensitivity analysis}\label{sec:out41}

The results of Section \ref{sec:diff-discr} suggest a weak role  of the latent variable $V$ on each exam enrollment variable $R_j$; see equation \eqref{eq:par2} for the specification of this relation. Indeed,  the discrimination parameters of $V$ ($\hat{\gamma}_{2j}^*$) in Table \ref{tab:discr} are significant (at $5\%$) for only one-third of the items, and they are lower in absolute value than the discrimination parameters of $U$  ($\hat{\gamma}_{1j}^*$). Moreover, the standard errors for the parameters of Mathematics courses are abnormally high (for details see standard errors  $\hat{\textrm{se}}_{{\gamma}_{Vj}^*}$  for group Q-Z in Table \ref{tab:discr} and also standard errors  $\hat{\textrm{se}}_{{\delta}_{j}^*}$  referred to all groups of Mathematics shown in the online Supplementary Material,Table 2).

To better assess the role of $V$, we compared
the selected model  ($k_U=4$,$k_V=2$ in Table \ref{tab:bic}) against a restricted  version without 
$V$ ($k_U=4$,$k_V=1$), entailing a reduction of model parameters from $226$ to $194$. In the restricted version of the model, the standard errors for Mathematics are no more problematic (see estimates shown in the online Supplementary Material, Tables 3-5). 
Moreover, the BIC index reduces from 14203.86 to 14166.19, thus confirming that  the importance of $V$ in terms of improvement of fit is small.
It is worth noting that the main findings about the effects of the latent ability $U$ are unchanged (online Supplementary Material, Tables 6-9).

Despite its small contribution to the model fit, the latent tendency $V$ gives additional insights into the student decision process, thus we decided to conduct the analysis using the model with both latent variables.
\section{Conclusions}
\label{sec:concl}

For the evaluation of student's proficiency we propose an Item Response Theory (IRT) approach that jointly account for the observed exam results (failed/passed exam and grade) and for the  information on the exam enrollment (at least once/never), whose absence gives rise to missing values about the exam result.  For this aim, we adopt a  multidimensional latent class IRT model, where the same latent variable (i.e., student ability) affects both the enrollment process and the exam result.  The proposed model follows a shared-parameter approach for the treatment of non-ignorable missingness and originates from the finite mixture structural equation model of Bacci and Bartolucci (2015), extended in a suitable way to account for ordered polytomous items and missing data indicators that are incompletely observed.  

The proposed model yields useful results for both students and administrators. 
In contrast to traditional approaches, our model explicitly considers the enrollment to single exams: this is crucial in our setting where first-year courses are compulsory, while order and timing of exams are chosen by the student.
Indeed, almost all students take only few of the exams during the first year, and thus they implement a strategy for choosing the order of the exams. It turns out that the enrollment rates to the exams are very different across courses, and in some cases also between groups of the same course. The analysis shows that the enrollment to exams depends on its perceived difficulty, as shown by differences in enrollment rates between and within courses. Moreover, we find that the enrollment does not significantly depend on student preferences (tendency to take exams in quantitative exams); on the other hand, it depends on the student ability, so that the enrollment mechanism is not ignorable. Therefore, the prediction of the student ability requires to jointly model  the enrollment decisions and exam results. 

We treat the exam result as an ordinal item with five categories: the first one standing for failing, and the remaining four categories for the grade assigned if the exam is passed. The predicted probabilities of passing the exams and of exam grades for a student with given ability show remarkable differences among disciplines. As  expected, Mathematics is the hardest exam, with low probability of passing the exam and low grades, highlighting problems with either the course content or the use of the grading scale.
Moreover, some courses show worrying differences between groups, posing a fairness issue. For example, for a student with average ability, the passing rate of Accounting ranges from 0.35 to 0.86, according to the group to which he/she is assigned, and for a student with average ability who passed the exam two groups of Microeconomics have the mode on the  grades 18-21, and  another one on the grades 25-27.
This fact poses a serious issue of fairness, given that the assignment of students to groups is based on surname, thus groups are expected to be homogeneous with respect to student ability. 

The proposed model allows us to cluster students into four latent classes of ability, corresponding to widely different performances.
The structural part of the model relates class membership to observed characteristics. The probability to belong to classes of greater ability is higher for students coming from a scientific high school, students with a good school grade, and students beginning university in the academic year following the end of high school.  This information can be used by potential freshmen and by the university management for planning guidance and tutoring activities.

To conclude, we outline that the proposed LC-IRT approach based on within-item dimensionality is suitable for a wide range of applicative problems  characterized by ordinal items with non-ignorable missing item responses, such as tests concerning the measurement of  customer satisfaction, quality of life, and levels of physical or psychological disabilities.    

\section*{Acknowledgements}
Authors acknowledge the financial support from the grant FIRB (``Futuro in ricerca'') 2012 on ``Mixture and latent variable models for causal inference and analysis of socio-economic data'', which is funded by the Italian Government (RBFR12SHVV).



\end{document}